\documentclass{goeproc}

\usepackage{shortvrb}
\MakeShortVerb{\|}
\def\pdfLaTeX{pdf\kern.06em\LaTeX}
\usepackage{bm}

\begin{document}

\title{Stokes Profile Inversion in Meso-Structured Magnetic Atmospheres}

\author[]{T.A.~Carroll}

\affil[]{Astrophysikalisches Institut Potsdam, Germany}
\affil[]{\textit{Email:} tcarroll@aip.de}

\runningtitle{MESMA}
\runningauthor{T.A.~Carroll}

\firstpage{1}

\maketitle

\begin{abstract}

Based on the Meso-Structured Magnetic Atmosphere (MESMA) approximation
\citep{Car06b} we present first results of an inversion of spectropolarimetric observations
obtained from internetwork regions. To cope with the inherent complexity of the
mostly unresolved magnetic field in the solar photosphere the MESMA approach 
provides a statistical description of the underlying atmosphere in terms of a 
random Markov field. This statistical model allows us to derive a 
stochastic transfer equation for polarized light. 
The stochastic transfer equation explicitly accounts for the
spatial correlation -- the characteristic length scale -- of the underlying magnetic and
non-magnetic structures. We use this new diagnostic parameter in an inversion approach 
to demonstrate that the magnetic flux structures in the solar internetwork possess a 
finite correlation length which is not compatible with the classical flux tube picture.

\end{abstract}

\section{Introduction}

  The entire solar photosphere exhibits a rich structure of large and small scale
  magnetic features like sunspots, pores  faculae or plages.
  But except for sunspots and pores these magnetic fields cannot be
  spatially resolved with present telescopes, although these fields
  clearly manifest themselves in high resolution spectropolarimetric observations.
  
  With the improvement of spectropolarimetric sensitivity and
  spatial resolution over the last years it became clear that
  these unresolved magnetic fields are much more ubiquitous than previously thought.
  This raises the question of the significance of these
  elusive and complex magnetic fields for the solar magnetism in general
  \citep{Schri03,SA04} and how these magnetic fields can be appropriately
  investigated by spectropolarimetric observations.
  
  The interpretation of Stokes profiles 
  in the context of the thin flux-tube model
  relies on the basic picture of an embedded cylindrical magnetic structure 
  surrounded by a quasi field-free medium.
  Based on that assumption a so called 1.5-dimensional radiative transfer is applied
  where a number of rays piercing through the underlying
  2- or 3-dimensional geometry of the model to obtain the 
  spectral 'signature' of the underlying magnetic structure \citep{Sol93}.
  But if the underlying structures are much more dynamic, disrupted and intermittent,
  the conventional static flux-tube model will allow only a poor representation 
  of the real magnetic field structure.
  In this sense the flux-tube modeling provides a rather macroscopic treatment of the problem
  -- in the 1.5 dimensional sense -- since the averaging process for of all 
  line-of-sights (LOS) is performed after the actual integration of the transfer equation.
  
  The other extreme, in contrast to the macroscopic view, is the
  MISMA approximation. The assumption here is that the atmospheric conditions along 
  the line-of-sight are rapidly changing. 
  The fluctuation of the atmospheric parameters occurs on very short scales, such that a
  micro-structured or micro-turbulent approach is justified. This allows an averaging 
  over all atmospheric parameters at each spatial position before the 
  actual transfer equation is integrated.
  Despite its appealing simplicity in the way this approach treats the
  radiative transfer the idealized assumptions about the underlying atmosphere
  strongly limits the application of this approach. Structures in the solar photosphere 
  whether magnetic or non-magnetic are in general not in a microturbulent state.
  Magneto-convective simulations suggests that neither predefined static macro-structures
  nor pure micro-structures are present in the solar photosphere, the possible structuring 
  seems much more to comprise a broad range of different scales. 
  \citep{Schaf05,Vog05,Stein06}.
  
  This paper is organized as follows: In Sect. 2 i briefly summarize the basic concept of 
  line formation in stochastic media and present the stochastic polarized transfer equation.
  In Sect. 3 i give an overlook of the first results of an inversion of spectropolarimetric
  observations obtained from internetwork regions. Sect. 4 concludes with a summary of the
  here presented analysis.

\section{The stochastic transfer equation for polarized light}
The approach described here is based on a statistical model of
the atmosphere in terms of a random Markov field, the
MEso-Structured Magnetic Atmosphere (MESMA) which was 
introduced by \citet{Car06b}. 
In this contribution i will just give a brief summary of the basic 
concept of the MESMA approach, for a more 
detailed presentation of the statistical model and derivation of the
stochastic transfer equation the reader referred to
\citet{Car03,Car05a} and \citet{Car06b}.

The atmospheric volume of interest (the actual resolution element) is 
assumed to be characterized by an a-priori unknown 
structuring along the line-of-sight. The only assumption we
make about the underlying atmosphere is that the
structures have a finite spatial extent and can be described 
in terms of a Markov random field.
This allows us to neglect all higher order spatial correlation effects 
to use a first order approximation to describe the spatial correlations.

We begin by introducing a random atmospheric vector $\bm{B}$ which
comprises all relevant atmospheric parameters such as temperature,
pressure, velocity, magnetic field strength, magnetic field
inclination, etc. If we move then along an arbitrary line-of-sight 
we obtain a series of realizations at different positions $s$ 
for the random vector $\bm{B}$.
This spatial dependency allows us to describe $\bm{B}$ in terms of a
stochastic process (a Markov process) along the line-of-sight and 
for which we can specify a suitable conditional probability density 
or transition probability from one spatial point $s$ to another $s+\Delta s$,
\begin{eqnarray}
p(\bm{B}_{s+\Delta s} \mid \bm{B}_s) = e^{\frac{\Delta s}{\lambda}} \:
\: \delta(\bm{B}_s - \bm{B}_{s+\Delta s}) \: 
+ \: (1 - e^{\frac{\Delta s}{\lambda}}) \: p(\bm{B}_{s+\Delta s}) \;. \label{kubo}
\end{eqnarray}
This conditional probability which specifies the so called 
Kubo-Anderson process \citep{Fri76} describes how the probability for a 
transition changes as we move along the line-of-sight from a given 
position $s$ and the associated atmospheric conditions 
$\bm{B}_s$ at this position.
The probability for staying in the same regime $\bm{B}_s$ for the 
entire trajectory $\Delta s$ decays exponentially 
while the probability for a sudden jump into another 
atmospheric regime $\bm{B}_{s+\Delta s}$ rapidly grows. 
The conditional probability density therefore describes the correlation 
of the individual structures between two spatial positions $s$ and $s+\Delta s$. 
The degree of correlation is controlled by the parameter $\lambda$, the characteristic length 
(correlation length) of the structures. Based on that particular Markov process
we can derive the following stochastic transport equation \citep[see][]{Car06b},
for the so called mean conditional Stokes vector $\bm{Y}$
\begin{eqnarray}
\frac{\partial \bm{Y}_B(s)}{\partial s} \: = \: - \bm{K \: Y_{B}} + \bm{j} \; 
+ \int\limits_{\bm{\hat{B}}} \lambda^{-1}_{B''}
\bm{Y_{B''}} \: p(\bm{B}'',s) \: d\bm{B}_{s}''  -
\int\limits_{\bm{\hat{B}}} \lambda^{-1}_B  \bm{Y_{B}} \: p(\bm{B}'',s) \: d\bm{B}_{s}''  ,
\label{stochstokes}
\end{eqnarray}
The mean conditional Stokes vector $\bm{Y}$ is defined as
\begin{equation}
\bm{Y}_{B}(s) \: = \: \int_{\bm{\hat{I}}} \bm{I} \: p(\bm{I},s \mid \bm{B},s) \: d\bm{I} \: .
\hspace{2.5cm}
\end{equation}
It is important to realize that $\bm{Y_{B}}$ is a statistical equation and it is 
conditioned on one particular atmospheric regime $\bm{B}$. The transport Eq.
(\ref{stochstokes}) describes the evolution of $\bm{Y_{B}}$
through the atmosphere along the LOS. 
There are four basic processes that govern the transport of $\bm{Y_{B}}$,
the two processes of absorption and emission and two more
processes that describe the statistical inflow and outflow of
intensity to and from the regime $\bm{B}$ under consideration.
The degree of statistical scattering and absorption is controlled 
by the correlation length $\lambda$  of the particular atmospheric 
structures. The observable of our problem -- the expectation value of the Stokes
vector -- at the top of the atmosphere $s_t$ can easily be obtained from a final
integration of the mean conditional Stokes vector over the entire
atmospheric state space $\bm{\hat{B}}$,
\begin{equation}
<\bm{I}(s_t)> \: = \: \int_{\bm{\hat{B}}} \bm{Y}_B(s_t) \: p(\bm{B},s_t) \: d\bm{B} \; .
\hspace{2.0cm}
\end{equation}
It is this extra degree of freedom in the stochastic transfer equation, given by the 
correlation length, which provides the additional diagnostic capability of the 
stochastic approach \citep{Car05b,Car06a,Car06b}. As could be shown by
\citet{Car06b} the asymmetry of the Stokes $V$ profiles 
directly depends upon the underlying correlation length and allows
to estimate the characteristic length scale of the
magnetic field from the net-circular-polarization (NCP) of 
Stokes $V$ profiles.

\section{Analysis of internetwork magnetic fields}
Based on the stochastic mesostructured approach we 
have analyzed observations of full Stokes profiles of the iron line 
pair Fe\,I at 630 nm, taken at the High Altitude Observatory/National Solar
Observatory Advanced Stokes Polarimeter (ASP). The quiet sun
data were obtained by B. Lites on 1994 September 29 \citep{Lites96}.
In the following we are in particular interested to gain some insight 
into the characteristic length scale of the magnetic structures in the 
internetwork.
\begin{figure}[h]
\center\includegraphics[width=0.75\linewidth]{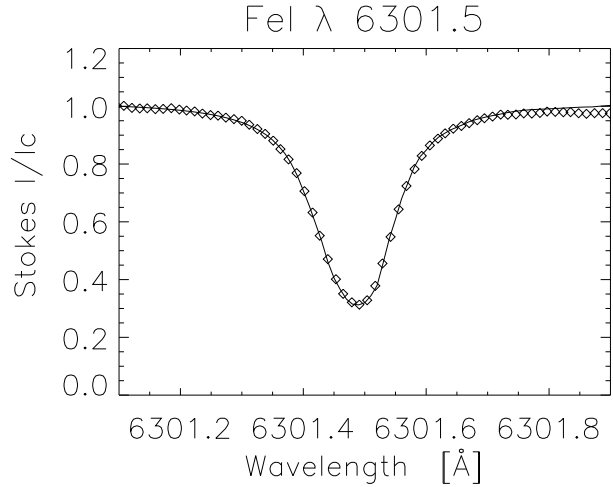}
\caption{A fitted Stokes $I$ profile of the Fe\,I 6301.5
nm line. Diamonds represent the observation, the solid line the 
profile fit.}
\label{fig:profiles1}
\end{figure}
Our inversion routine is based on the Levenberg-Marquardt algorithm \citep{Press92}
and incorporates the stochastic transport equation as the forward kernel.
In a first step we analyzed granular and intergranular regions in
order to determine the convective characteristics of the
non-magnetic components from Stokes $I$ profiles. 

For the granular and intergranular
components we adopted the granular and intergranular model
atmospheres from \citet{Bor02}. Based on the
temperature and pressure stratification of these models we 
assumed a simple 3-type stochastic
velocity field. Please note, that the stochastic approach makes no 
assumption about the numbers of individual structures in the resolution element, 
the only assumption made here is that there are three different types of structures
present.

The free parameter of the fitting routine are the
three single-valued velocities for each atmospheric component, 
the probability values of the individual types (comparable to the
conventional filling factor) and the correlation lengths of the
individual components. For an upflow (granular) region a fit of a
Stokes $I$ profile is shown in Fig.(\ref{fig:profiles1}). 
The remarkable result here is the fact that there is no need to 
use nonphysical parameters like micro- or macro-turbulence to fit the 
profiles. The convective velocity structure exhibits a clear 
mesostructured behavior on scales between 
200 km and 400 km.
\begin{figure}[h]
\center\includegraphics[width=0.75\linewidth]{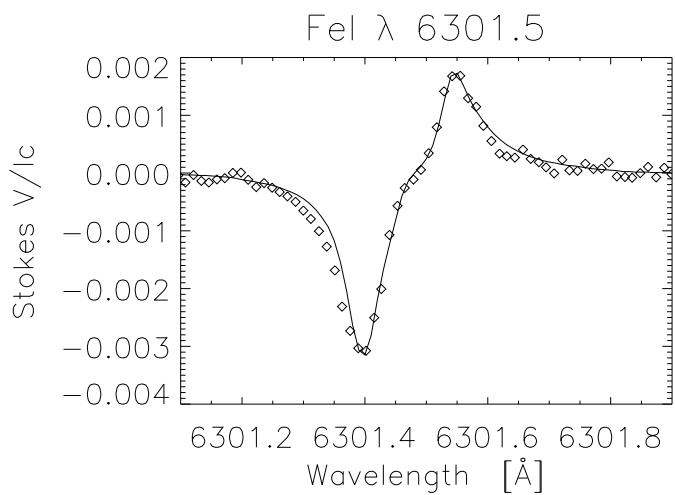}
\caption{Fit of a typical Stokes $V$ profile of the Fe\,I 6301.5
nm line, again the diamonds represent the observation and 
the solid line the fit to the profile.}
\label{fig:profiles2}
\end{figure}
\begin{figure}[h]
\center\includegraphics[width=0.75\linewidth]{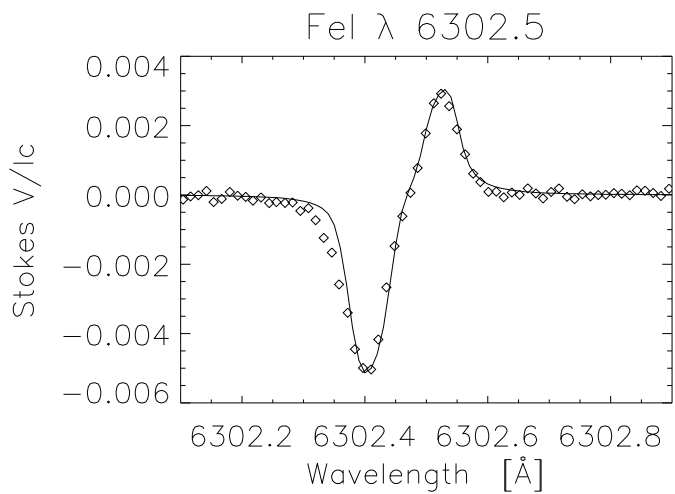}
\caption{ A typical fit for the Stokes $V$ profile of the Fe\,I 6302.5 nm line.
Diamonds observations, solid line fit.}
\label{fig:profiles3}
\end{figure}
These results are in agreement with earlier investigations
of mesoturbulent velocity fields in the solar photosphere by \citet{Gai76}. 

As we could place tight constraints upon the characteristics of the ambient
flow pattern we proceed by analyzing the magnetic field 
by inverting the respective Stokes $V$ profiles. 
We adopted a stochastic model of two different types of structures 
(non-magnetic and magnetic). The free parameter 
of the inversion are the velocities, magnetic field strengths 
as well as the correlation lengths of the individual structures. 
Two fits of the Stokes $V$ profiles are
shown in Fig.(\ref{fig:profiles1}) and Fig.(\ref{fig:profiles2}) . 
The observed Stokes $V$ profiles could be well reproduce by the 
stochastic two-ensemble model. 
In particular the asymmetries of the Stokes
$V$ profiles are well reproduced.
The magnetic structures in the analyzed internetwork region have surprisingly 
small correlation lengths between 50 km and 125 km for structures in upflow elements 
and 150 km to 230 km for structures in downflow elements.
These results clearly indicate that the magnetic field -- as well as the
velocity field -- in the internetwork can neither by described 
by macroscopic structures like flux tubes nor can they be 
described in terms of a microstructured or microturbulent field. 
A interesting result here is the clear trend for stronger field 
structures to have larger correlation lengths and for weaker structures 
to have shorter correlation lengths. 
This seems to be the result of the increased buoyancy forces 
of strong magnetic structures which results in a preferred vertical 
alignment and gives rise  -- for disc center
observations -- to an increase of the line-of-sight correlation lengths.
 
\section{Summary}
This analysis of magnetic field structures 
in the internetwork clearly demonstrate the feasibility of 
an inversion under the MESMA concept. Moreover, we found that
the characteristic length scales in the solar internetwork 
are relatively small, but clearly far from being  
microstructured or microturbulent.
However, the length scales found in this work are also not
consistent with the classical flux tube picture \citep{Stenflo94}
which would require correlation lengths larger than 350 km.
Another intriguing result here is the fact that the 
obtained correlation lengths of the magnetic structures 
agrees very well with the autocorrelation (along a vertical direction) 
of magnetic structures in magnetohydrodynamic simulations
\citep{Schaf05} which have a mean value of approximately 
220 km. This surely deserves further investigations but one can
already say that advanced magnetoconvective simulations and (polarized) radiative 
transfer modeling provide a fruitful combination to gain further insight into 
the surface magnetism of the sun.

\begin{acknowledgements}

We gratefully acknowledge support of this work by the Deutsche Forschungsgemeinschaft (DFG) under the grant
CA 475/1-1.

\end{acknowledgements}


\end{document}